# Multiscale Biophysical Waves (MBW): Conceptual and Theoretical Framework

Samina S. Masood, Alishpa Masood, and Robert L. Jones Jr.
Mathematical and Physical Sciences
University of Houston, Clear Lake

## Abstract

This paper establishes a conceptual and theoretical framework for multilevel communication between quantum, molecular, cellular, tissue-organ, whole body, and other biophysical spaces. The wave-mechanical description of electromagnetic signaling is developed, detailing how waves move through cellular structures and interact energetically with surrounding biomaterials. These regions transmit and receive partially coherent biomolecular signals within and across cells, tissues, organs, and neural networks, where resonant frequencies interfere, converge and respond. These processes introduce nonlinearities and stochastic delays that shape timing, coherence, and network-level dynamics, and can be integrated within a pragmatic framework linking mathematical modeling of the physical state to experiential outcomes. This contribution stands as an independent theoretical framework: it sets the foundation for the wave mechanical approach to understanding brain function, which may help to develop new methods to prevent aberrant brain behavior. Mathematical derivations of the inter-sector coupling operators and clinical and therapeutic applications are postponed for future work.

## 1: Introduction

Communication plays an integral role in coordinating biological behavior within living systems and between complex organisms. More specifically, intercellular transmissions propagate through various pathways and mechanisms to achieve signal conduction and transduction, which are highly consequential for homeostasis, development, and a myriad biophysical processes.[1] Despite technological advancements that have made it possible to record communication coordinated across multiple sites, there is no consensus on a unified framework linking quantum activity, such as biophoton emission to cellular signaling and higher functions in biosystems.[2]

Living creatures communicate in and among themselves using a wide range of sensory modalities, signaling strategies, and transmission and reception resources. Even the smallest mammals, birds, insects, plants, and microorganisms possess specialized mechanisms to exchange information and sense their environment.[3,4] In the human nervous system, neural oscillations have been implicated in sensory, motor, and cognitive processing.[5] At the biochemical level, communication manifests primarily as chemical, mechanical, and ionic signaling modes, with associated local electromagnetic fields rather than as freely propagating electromagnetic radiation. However, at the most fundamental physical level, biophysical signaling arises from electromagnetic interactions among charged particles and dipoles.[6,7]

At the quantum scale, atoms and molecules exhibit wavefunctions and quantized vibrational modes that provide a microscopic basis for biophysical wave phenomena. Interactions among electrons and nuclei determine the structure, reactivity, and dynamics of molecular systems.[8] These quantum-level oscillatory processes form the foundational tier of biological organization, propagating upward through biomolecules, cells, and organ systems.[9,10] Conversely, higher-order biological structures impose top-down constraints that modulate quantum and molecular dynamics, producing a bidirectional interplay across various scales.[11] This reciprocal organization, where macroscopic biological states shape microscopic processes and vice versa, aligns with contemporary multiscale frameworks in molecular biology and micro- and macro-level systems in biophysics, which emphasize hierarchical coherence, cross-level coupling, and emergent electromagnetic signaling in living systems. This analysis of inter- and intra-biophysical communication is conducted in terms of wave mechanical concepts such as dissonance, consonance, interference, coherence, and resonance.

In the next section the various contributions of biomechanics are discussed in detail. Section 3 is comprised of the Mathematical formulation of wave mechanics of brain which is responsible for communication between brain and rest of the body. In section 4, we describe the communication system in living organisms in terms of multiscale biophysical waves (MBW), developing a bridging system between various components from molecular level to the entire body, which communicates with outside the world as well. Section 5 is reserved to discuss the possible application of this integrated approach of MBW communication.

## 2. Biochemical Physics of Brain

As biological systems increase complexity, multicellular organisms can generate structured oscillatory patterns in which constructive interference enhances amplitude and enables modulation of biophysical signals. Synchronization of wave interactions across increasing levels of biophysical organization gives rise to distinct forms of wave mechanics that support communication and control. These modulated field patterns can influence neural activity and contribute to the regulation of cellular processes, shaping or encoding biological responses as they propagate through biological media and transmission pathways.

Chemical energy converted into pulsatile and oscillatory electrical activity is transmitted through biochemical, physiological, cellular, and tissue-level processes as signals propagate through elements of the component networks.[12] As these signals travel, they are continually modified by the surrounding biological medium and by local electric and magnetic fields, which shape their amplitude, timing, and coherence.

Radiation interacting with matter undergoes velocity reduction and dispersion due to the intrinsic electromagnetic properties of the medium. Analogously, oscillatory signals in biological systems are shaped by the heterogeneous and frequency-dependent characteristics of biological tissue. Electromagnetic signals are influenced by variations in thermal vibration, electrochemical potential, chemical composition, and mass distribution, all of which affect propagation energies. A full understanding of radiation–matter interactions require a framework of quantum electrodynamics (QED), which incorporates the interconversion of energy and matter in four-dimensional relativistic analysis. QED can be considered as a four-dimensional (relativistic) generalization of low energy quantum mechanics and represents classical electrodynamics at macroscopic level however; electrodynamics is the only fundamental interaction which is well-understood at individual particles level (as small as electrons), to macroscopic electrodynamics.

Nonrelativistic quantum mechanical limits describe individual atoms, molecules, and chemical bonding; the same electromagnetic principles governing atomic and molecular structure are essential for understanding cellular chemistry. Relativistic electrodynamics and quantum electrodynamics (QED) provide the universal theoretical framework for electromagnetic interactions, while nonrelativistic quantum mechanics and classical electrodynamics supply the effective descriptions appropriate to biological energy scales. These form a complementary hierarchy: the lower energy biological models are consistent reductions of the more general relativistic QED formalisms. QED links chemically generated particle energies with electromagnetic interactions and describes how wave behavior is modified when fields couple to matter.

Electromagnetic interactions at the quantum scale give rise to oscillatory behavior that can influence low-energy signaling through interference at molecular and sub-molecular levels. When electromagnetic fields couple to charged particles in matter, they form matter-dependent modes that differ from free-space radiation and can exhibit longitudinal components due to interactions with the medium.[13] These interactions of electromagnetic fields with matter shape neural resonance, constructive and destructive interference, and energy-dependent propagation characteristics within biological tissue.[14,15] Integrated effect of the interaction of radiation with matter is observed in terms of refractive index in optics which is related to the nature of the medium described by its electromagnetic behavior. However, its explicit derivation is possible in QED and its classical limit[16].

Biochemical processes arise from the quantum-mechanical behavior of atoms, whose electron distributions (due to Fermi statistics) generate electrical potentials that govern bonding, molecular conformations, and the interactions underlying biological behavior.[17] Electromagnetic interactions at the molecular scale produce oscillatory fields that can superimpose with neural carrier signals, contributing to the composite patterns that propagate through neural pathways.[18,19]. As these signals travel, they are distributed across brain regions according to their frequency content and amplitude. Neural structures can modulate amplitude, frequency, or phase to refine transmission, enabling selective activation of the corresponding functional neural networks.

Biomolecules generate and interact with weak electromagnetic fields arising from ionic currents and the vibrational modes of cytoskeletal structures. At the macroscopic scale, synchronized ionic currents across neuronal populations produce electromagnetic field patterns responsible for neural tissue vibrations, which are detected as brainwaves. These oscillations coordinate perception, cognition, and motor control through established neural pathways and contribute to the temporal organization of large-scale functional networks.[20,21,22]

Nervous tissues, being an integral part of the neural network, provide control and communication that regulates interactions across the body.[23,24] Neurons also contain nucleic acids and aromatic amino acids whose electronic structures support molecular recognition, stability, and information transfer. Their configurations and the ability to produce energy influence electromagnetic oscillations that regulate cellular growth and death and determine their response and effect on behavior. These electrodynamic processes form the foundation of neural signaling and contribute to the emergence of coherent brain activity and may help to understand brain functionality and consciousness.[25]

In vertebrates, the brain orchestrates information transfer throughout the organism by integrating diverse signaling processes, including synaptic transmissions, action potentials, endocrine signaling, autonomic, and immune cytokine signaling, with larger-scale electromagnetic oscillations (brain waves), providing a temporal and spatial framework that modulates, synchronizes, constrains, and generates physiology.

At the organ level, the brain is one of the most complex structures in nature. It sustains the organism by coordinating multiple independent regulatory systems that perform mutually dependent physiological functions, including the integration of signals originating both inside and outside the body. Functionally, the brain can be described in terms of two major divisions: the cortex, responsible for higher-order integration, representation, and voluntary control; and the subcortex, responsible for autonomic regulation, affective processing, and rapid sensorimotor responses. Together, these divisions form a coupled hierarchy that links deliberate cortical processing with fast, evolutionarily conserved subcortical mechanisms.

At the tissue-organ level, functionally unified multicellular assemblies, such as the heart, brain, and skeletal muscle, generate collective oscillatory fields whose coherence extends beyond any single cell. These structures act as organized wave sources: the heart produces rhythmic cardiac waves detectable as electrocardiographic (ECG) signals; neuronal populations of the brain generate coordinated rhythms measurable as electroencephalographic (EEG) oscillations; and skeletal muscle produces myoelectric waves (i.e., electromyography, or EMG) arising from synchronized motor unit firing. At this scale, the relevant oscillatory modes are emergent properties of multicellular architecture rather than properties of individual cells, and their electromagnetic signatures propagate through surrounding tissue.

Electromagnetic fields arising from ionic currents and charge separation in neural tissue obey basic principles of electrodynamics (Maxwell's equations) and provide a classical electrodynamic description of brain activity. These fields, generated by biochemical and electrochemical processes across membranes and synapses, form low-frequency, quasi-static electromagnetic signals rather than freely propagating waves. Their behavior is dominated by near-field, diffusive electrodynamics due to the conductive and dielectric properties of neural tissue, which prevent the formation of radiative electromagnetic waves at physiological frequencies. Their spatial and temporal structure reflects the underlying neural dynamics and enables macroscopic measurements such as EEG and EMG.

Electromagnetic waves generated by biological tissues undergo amplitude and frequency modulation and transverse–longitudinal decomposition in matter, as the common framework for the series (atomic, molecular, cellular, tissue/organ. whole-body. environment). It also develops the near-field evanescent formalism explicitly, including the evanescent field equation, the transverse–longitudinal mode decomposition, the superposition of modes at biological interfaces, and the overlap integral: together these constitute an original wave-mechanical treatment of biological interface coupling grounded in established electrodynamics.

## 2.1: Sources of Brain Waves

Brain waves emerge from the coordinated generation of oscillatory electromagnetic fields produced by synaptic currents, membrane potential fluctuations, and population-level firing dynamics. At a mesoscopic scale, these fields arise in cortical circuits through the collective electrochemical activity of cell assemblies which facilitate interareal communication across distal brain regions through hierarchical processing between cortical and subcortical structures.[19] These neural population dynamics emerge in the greater limbic system with a combination of electrical and chemical signaling involved in synaptic transmission that is routed through excitatory and inhibitory pathways. Within a single neuron, action potentials may be generated spontaneously via ongoing recurrent activity or evoked by an external stimulus (i.e., sensory receptors) transmit electrical waves that are conducted by myelinated axons for fast processing. Chemical signaling is carried out by neurotransmitters such as glutamate, GABA, and acetylcholine regulate the excitatory–inhibitory balance that shapes these rhythms, while the structural and biophysical properties of neurons and glia determine the spatial and temporal coherence of the resulting oscillations. Brain waves thus represent a multiscale electrodynamic phenomenon, integrating microscopic ionic fluxes with mesoscopic circuit dynamics and macroscopic field propagation across the brain.

Canonically coordinated electromagnetic waves in the brain are classified into gamma (30–100 Hz), beta (13–30 Hz), alpha (8–13 Hz), theta (4–8 Hz), and delta (0.5–4 Hz) bands, each associated with distinct temporal structures and functional roles. These oscillations originate from specific laminar structures within the neocortex [26] and support communication across cortical and subcortical regions.

Low frequency neural oscillations have long effective wavelengths, but in biological tissue they manifest as quasi-static, near-field electric and magnetic fluctuations whose spatial reach is limited by the conductive and dielectric properties of the medium. The transverse electromagnetic components remain confined to reactive near-field domains, where they influence dendrites, adjacent neurons, local microcircuits, and the surrounding extracellular ionic environment.[27]

**2.2 Electrochemistry of the Brain**

It is not realistic to describe the chemistry of every neuron nor to track the full spectrum of electromagnetic waves generated by each biochemical event where this number of interacting degrees of freedom provides justification to apply statistical analysis. In broad outline, the oscillatory behavior of neuronal assemblies follows from the configuration of nucleic and amino-acid polymers and the supramolecular structures they form. The aromatic amino acids arranged within subcellular architectural structures such as the microtubule interior form channels whose polarizable electron clouds generate electromagnetic field gradients and sustain oscillatory modes that propagate as electrodynamic waves[28,29,30.]

Aromatic residues such as phenylalanine, tyrosine, and tryptophan, along with heterocyclic rings (histidine and the purine and pyrimidine bases of nucleic acids), can adopt conformations that generate localized electromechanical oscillations theoretically responsible for cellular and sub-cellular communication[31,32]. These molecular configurations create waves which converge, interfere, and synchronize through resonance channels, forming synchronous superpositions. These superpositions have been proposed in theoretical models to contribute to higher-order integrative processes, including aspects of consciousness.

Depending on the biophysical predisposition of the neuron, an incoming wave may transmit, reflect, or undergo interference, shaping the resulting electrodynamic pattern while remaining consistent with the principles of electrodynamics and quantum mechanics.

Incoming waves can excite or de-excite atoms, generating characteristic superposition patterns. These wave patterns may interfere or undergo modulation, activating selected neural regions after traversing the brain due to the resonance phenomenon. The brain may simply ignore low-intensity, low-energy waves as perturbative vibrations. These vibrations may gather as coherent signals and interfere to become more intense signals or undergo the frequency changes due to modulations. These waves then propagate through the surroundings and to areas determined by the resonant characteristics of specific brain regions. The mathematical formulation developed in Section 3 describes the scope and dynamics of these waves and their role in brain function. A complete understanding of this mechanism could open new avenues for neural treatment and suggest more effective drugs with minimal side effects.

## 3: Mathematical Formulation

Waves are represented as a wavefunction (or state function) $\psi$, which may represent electromagnetic signals and/or matter waves. They are chemically generated by multicellular systems and are described mathematically as superimposed wavefunctions $\psi = \Sigma\psi_i$, where index i corresponds to various frequencies f, and the energy of each (`i'-th) wave is given by $E_i = hf_i$, where h is Planck's constant ($\approx 6.63 \times 10^{-34}$ J·s). Below we define Multiscale Biophysical Waves (MBW) in a more formal way to correlate brain functionality and the wave mechanics.

Living cells are composed of complex organic molecules including carbohydrates, lipids, proteins, and nucleic acids, whose structures arise from the electromagnetic interactions among charged and polarizable atomic constituents. These molecules possess permanent and induced dipole moments, and their collective electrostatic and electrodynamic properties shape biochemical reactivity, molecular recognition, and higher-order cellular organization. Electrochemical energy generated by cellular reactions is determined by electromagnetic waves using Maxwell's equations, which combine electromagnetic wave behavior with the electromagnetism of charges. Accepting the absence of magnetic monopoles, the charges obey the fundamental laws of electromagnetism namely Gauss's law, Faraday's Law, Ampere's law. Differential forms of Maxwell's equations in free space (SI units) is written as:

$$\nabla \cdot B = 0 \qquad (1)$$

$$\nabla \cdot E = \rho/\varepsilon_0 \qquad (2)$$

$$\nabla \times E = -\partial B/\partial t \qquad (3)$$

$$\nabla \times B = \mu_0(J + \varepsilon_0\, \partial E/\partial t) \qquad (4)$$

where $\varepsilon_0$ and $\mu_0$ are the electric permittivity and magnetic permeability of free space, and $c = 1/\sqrt{(\varepsilon_0\mu_0)}$ is the speed of light. $\nabla \cdot E$ describes how electric field lines emanate from charge sources, while $\nabla \cdot B = 0$ confirms the absence of magnetic monopoles. The magnetic field lines form closed loops with no sources or sinks. $\nabla \times E$ and $\nabla \times B$ capture the rotational (swirling) nature of electromagnetic fields and relate them to electromagnetic waves. When these waves pass through matter, they interact with the medium and undergo changes in propagation speed, described by the complex relative permittivity $\varepsilon_r$ and relative permeability $\mu_r$ of the tissue.

Our earlier study of calculations for electric permittivity, magnetic permeability, and refractive index at relativistic energies (Masood et al., 2019)[33] may not appear directly aligned with neurobiophysical applications however, they rigorously characterize the conditions under which longitudinal electromagnetic modes arise: modes whose existence and dynamics are governed by charge-density fluctuations and medium-dependent field constraints rather than energy alone, unlike transverse waves. It incorporates the interaction of electromagnetic signals with the medium of transmission, and this interaction develops into longitudinal component of the waves which propagate through the tissues. The key parameter of studying the interaction with the medium is the significant ratio between thermal and chemical energy, which remains equally important and remains relevant even at the high-density classical limit.

The dependence of an electromagnetic wavefunction as $\psi(x,t)$ on space (**x**) and time (**t**) is written as

$$\psi(x,t) = e^{(ikx - i\omega t)},$$

where $\omega = 2\pi f$ is the angular frequency in radians per second, and f is the frequency in Hertz. The wavefunction has two components: a transverse component and a longitudinal component.

Electromagnetic waves are transverse in free space, but in a medium they may acquire a longitudinal component (as shown in Masood[33] and Masood & Saleem[14] and references therein).

### 3.1 Mathematical Representation

When an electromagnetic wave encounters a biological interface, the boundary conditions prevent full radiative transmission and force the normal component of the wavevector to become imaginary, producing an evanescent mode.

$$E_{ev}(x,z,t) = E_0 e^{-\kappa} \; e^{i(k_x x - \omega t)}, \qquad \kappa = \sqrt{k_x^2 - k^2} \qquad (5)$$

For an incident wave with in-plane wavevector $k_x$ exceeding the medium's wave number ($k = 2\pi/\lambda$) inversely proportional to the wavelength ($\lambda$), the evanescent field takes the (damping) form (eq. 5) [34], where the decay constant κ governs penetration depth. Larger κ corresponds to more tightly bound, rapidly decaying modes:

$$E_{ev} = E_T + E_L, \qquad \nabla \cdot E_T = 0, \qquad \nabla \times E_L = 0 \qquad (6)$$

The field admits a transverse–longitudinal decomposition (eq. 6), where $E_T$ is the divergence-free transverse component and $E_L$ is the curl-free longitudinal component. Evanescent modes always include a longitudinal component, distinguishing them fundamentally from free-space radiation:

$$E_{tot} = E_{inc} + E_{refl} + E_{ev} + E_{ind} \qquad (7)$$

At a biological interface Ω, the total field is a superposition of all contributing modes (eq. 7). The total electrodynamic environment experienced by neural and molecular structures is therefore not a single homogeneous field, but an interference pattern shaped by the overlap of externally driven and internally generated excitations[35]: where $E_{inc}$ is the incident transverse wave, $E_{refl}$ the reflected component, $E_{ev}$ the evanescent mode, and $E_{ind}$ the matter-induced longitudinal and polaritonic modes generated by charge motion and molecular polarization in the medium.

The coupling strength between any two modes $\psi_i$ and $\phi_j$ across the interface region is quantified by the overlap integral (eq. 8):

$$\langle \psi_i, \phi_j \rangle = \int_\Omega \psi_i^*(r)\, \phi_j(r)\, d^3r \qquad (8)$$

Equation 8 represents the interaction or transition between two waves ($\psi$ and $\phi$)**:** the coupling strength between two biophysical modes, quantifying how efficiently the incident mode $\psi_i$ can transfer energy, information, or coherence into the target mode $\phi_j$ through their spatial overlap within the interface region**.** This transition could be between longitudinal and transverse components or between electromagnetic signals and carrier waves depending on the point of interaction.

**4: Mechanics of Brain Wave Propagation**

Brain waves are effectively transmitted as carrier waves, transferring information between the brain, the rest of the body, and the external environment. These waves carry electromagnetic frequencies across the neural network, encompassing all energy waves produced by chemical reactions or received from the environment as transverse light waves, longitudinal sound waves, or other forms of carrier waves. Low-energy brain waves undergo interference, diffraction, and modulation while interacting with certain regions and bypassing others by resonance as they propagate through the neural network.

Resonant signals with matching characteristic frequencies of neurons play an effective role. The total wavefunction $\psi$ representing k superimposed waves is written as a linear combination of component wavefunctions:

$$\psi = \psi_1 + \psi_2 + \psi_3 + \ldots + \psi_k$$

where each $\psi_i$ corresponds to a wave of characteristic frequency $f_i$, with associated energy

$E_i = hf_i$. In principle, this superposition can be decomposed into its constituent frequency components, analogous to spectral analysis, giving a summation of the longitudinal and transverse components separately.

Superposition is distinct from modulation: modulated waves carry either a shifted frequency (FM) or altered power (AM) and cannot be simply separated, whereas superimposed waves retain their individual identities within the combined wavefunction.

**4.1 Modulation of Waves**

Modulation is a mechanism by which a carrier wave is systematically varied in amplitude (AM) or frequency (FM) to encode and transmit information. The AM wave is represented as:

$$m(t) = A_{m}cos(\omega_{m}t)$$

The carrier signal is:

$$c(t) = Accos(\omega ct)$$

The modified (AM) carrier amplitude is:

$$y(t) = Ac(1 + A_{m}/Ac)cos(\omega ct)$$

where ($A_m/Ac$) is the modulation index. If this index is too low, the modulation has no significant impact, and neuronal vibrations are damped out automatically. Larger values may be damaging, with the scale determined by the modulation index magnitude.

An FM signal can be represented as:

$$s(t) = sin(\omega ct + kAsin\omega_{l}t)$$

where $\omega c$ is the carrier angular frequency, $\omega_l$ is the locally generated angular frequency of the oscillatory field in the neural microcircuit being modeled, $k$ is the frequency sensitivity constant or frequency deviation constant in rad/s per unit amplitude, and $\beta = \Delta\omega/\omega_{l}$ is the FM modulation index, with $\Delta\omega$ the peak frequency deviation. An uncontrolled or pathological increase in power can cause brain damage and cognitive impairment (mild or severe). Signal strength may also diminish with age because of declining cellular health [36] It has been proposed that acoustically generated longitudinal waves may interact with neural tissue via mechanisms related to the brain wave framework described here, a hypothesis that requires further experimental validation[37],[38].

The formalism developed above provides the quantitative underpinning for the response wave dynamics. The superposition framework accounts for how low-intensity perturbative vibrations accumulate into coherent signals, while the modulation framework explains how those signals are encoded, amplified, or attenuated as they propagate to resonant brain regions. Together, AM and FM modulation indices constrain which response waves achieve sufficient amplitude to activate downstream neural circuits, and which are damped out, directly linking the mathematical parameters.

In the next section we define the Multiscale Biophysical Waves in a more formal way to correlate brain functionality and the wave mechanics.

**5: Multiscale Biophysical Waves**

In this paper, we develop a comprehensive framework for brain function based on the multiscale behavior of interchangeable wave forms, which we call *multiscale biophysical waves* (MBW). For the purpose of this work, we consider three scales: microscale, mesoscale, and macroscale. The microscale typically describes molecular and cellular activity incorporating the quantum mechanical atomic behavior; within the central nervous system (CNS). This broadly applies to single neurons and synaptic transmission (along with their respective subcellular structures). The mesoscale is particularly useful in the field of neuroscience as it graduates from individual cells to cell assemblies (i.e., neuronal ensembles) and describes the localized, collective activity of these specialized microcircuits of neurons that can communicate across brain regions. Finally, at the macroscale, whole-brain dynamics emerge from the functional connectivity and coordination of larger neural networks that facilitate sensory, motor, and cognitive processing. The benefit of a multiscale approach is it enables us to elucidate the role of wave behavior in hierarchical processing that may bridge the gap between molecular mechanisms and the network dynamics that contribute to higher brain function. These waves are responsible for the transfer of energy regardless of being longitudinal (carrier waves) or transverse (electromagnetic waves). These waves continuously transform energy as they propagate into, out of, and within the brain; they can also modulate or restructure other waves encountered along their trajectories. Moreover, they can acquire a longitudinal or transverse nature according to the requirement of the medium of propagation.

This wave-mechanical perspective provides a unified language for describing how electromagnetic signals are generated, how they are converted into mechanical carrier waves, and how wave phenomena at multiple spatial and temporal scales mediate energy transfer across neural structures. Consequently, the mechanical and functional properties of the brain can be interpreted through the specific wave types generated or transmitted along a particular segment of a neural pathway.

A biophysical wave represented by a wavefunction $\Psi_{BP}$ is a more detailed representation of the same longitudinal/transverse combination of waves represented earlier by Equation 8. We now write the biophysical wave as

$$\Psi_{BP} = \Psi_{BP}\,[E(x,t), B(x,t), u(x,t), \rho(x,t), \phi(x,t)] \qquad (9)$$

where E(x,t), and B(x,t) represent the time dependent electric and magnetic fields, u(x, t) is the displacement whereas ρ(x,t), correspond to density (mass or charge density), and ϕ(x,t) is the corresponding potential (gravitational or electromagnetic) at a point x, at a given time t. This relationship defines a generalized wavefunction $\Psi_{BP}(x, t)$ which simultaneously works for electromagnetic and mechanical waves.

At this point, it is necessary to define multiparticle-generated electromagnetic forces in a field-theoretic framework. In quantum electrodynamics, the appropriate construct is the Maxwell stress tensor (**MST**), which describes the local distribution of electromagnetic momentum and the tensile or compressive stresses exerted by the field at every point in space. It is written as

$$\sigma_{ij} = \epsilon E_i E_j + \frac{1}{\mu} B_i B_j - \frac{1}{2}\left(\epsilon E^2 + \frac{1}{\mu} B^2\right)\delta_{ij} \qquad (10)$$

where electric permittivity $\epsilon$ and magnetic permeability $\mu$ are the properties of a medium. The tensile stress $\sigma_{ij}$ represent the mechanical force produced by electromagnetic force responsible for creating mechanical vibrations (or carrier waves) in a medium. It transmits energy through cells and tissues, or continue to the next connected unit of the neural network as electromagnetic signal. These interchangeable electrical and mechanical waves can generate pressure to deform soft tissues or convert into pressure waves for fluids. These signals can incorporate the change in electric permittivity, magnetic permeability of a medium, or the propagation velocity which describe the interaction of radiation with matter.[14] The electromagnetic parameters of the medium describe the modification in electromagnetic signals, their propagation energy, and the change in direction of propagation as described by refraction of electromagnetic waves in a medium. This will also modify the MST in a medium,

This new MST interacts with the regions of neural network depending on the chemically created electromagnetic properties such as membrane capacitance (charge per unit potential). Therefore, the information transfer through a neural network can be described as a multiscale biophysical wave generated at every component of neural network based on the electromagnetic behavior of individual component which depends on its chemical compositions. It means that the wavefunction $\Psi_{BP}$ will acquire various forms (electromagnetic or mechanical) which could be purely transverse, longitudinal or vector waves with both components (electromagnetic waves with a longitudinal component). These multiparticle systems produce a combination of simultaneously produced waves which can undergo frequency modulation or amplitude modulation, where the microtubules or neural cells act as modulators. The simultaneous production of waves is responsible for the temporal coherence, which can facilitate interference as well. However, a further detailed study is required to fully understand the role and limitations of various components of neural network as frequency modulator or amplitude modulator.

### 5.1 Near-Field and Far-Field Electromagnetic Modes in Biological Media

The propagation of brain signals is a complex biophysical process. Neural communication relies primarily on electrochemical signaling, and these signals can be influenced by the structural and functional properties of neural tissue. Variations in membrane composition, ion-channel distribution, and network connectivity can modulate how signals are amplified, attenuated, or transformed as they travel through neural circuits.

When these modulatory processes become disrupted through altered excitability, impaired synaptic transmission, or abnormal network synchronization, the information may be misrouted, distorted, or delivered with inappropriate intensity. Such disruptions can lead to atypical patterns of neural activity, which may manifest behavioral or cognitive abnormalities.

Therefore, electromagnetic communication across biological interfaces cannot be captured adequately by purely particle based or static field descriptions, because the relevant processes involve propagating and evanescent wave modes that arise when electromagnetic fields interact with charged and polarized matter. In free space, propagating electromagnetic waves are strictly transverse, with electric and magnetic fields orthogonal to the direction of propagation. Whereas within biological media oscillating charges, dipoles, and polarization currents generate nearfield modes that include both transverse and longitudinal components and are strongly shaped by the dielectric and conductive properties of the tissue.[39] We may call them mechanical waves as the transfer energy through the biological media and their frequency and amplitude may vary with distance as well. Mechanical carrier waves are generated at an interface, The incident transverse field induces energy for the motion of charge and /or polarize ethe medium, which in turn generates matter coupled modes (including longitudinal and mixed polaritonic excitations) that remain bound to the material and decay rapidly with distance. [40] The total field in the vicinity of the interface is therefore a superposition of the incoming radiative wave and the locally generated nearfield modes, producing interference patterns that modify amplitude, phase, and frequency-dependent attenuation.[18] In biological tissue, synchronized ionic currents and molecular dipoles contribute to these nearfield components, so that the electrodynamic environment experienced by neural and molecular structures reflects the overlap of externally driven and internally generated wave modes, rather than a single homogeneous field. Brain waves, as measured by EEG, represent the volume-conducted potentials of these superimposed near-field modes, and their interpretation therefore requires accounting for the mode structure at the source, not merely the propagating signal at the detector. This inherently wave mechanical picture—superposition, interference, and mode conversion at interfaces—motivates a Hilbert space description in which distinct classes of electromagnetic excitations are represented as overlapping subspaces with well-defined inner products and coupling operators — that is, a mathematical framework in which different wave modes are treated as coexisting, interacting fields rather than independent signals.

### 5.2 Evanescent Modes and Interface Coupling in Biological Media

Evanescence occurs when a wave meets a boundary that cannot sustain its transverse component, producing an imaginary normal wavevector and exponential decay into the medium. The evanescence phenomenon indicates that the wave does not transfer energy without making net changes in the waveforms. The total internal reflection of electromagnetic signals may result in evanescence as well which is not due to opacity.

Evanescent electromagnetic modes arise when a wave encounters a boundary at which the wavevector component normal to the interface becomes imaginary. Rather than propagating into the second medium, the resulting field decays exponentially in perpendicular direction as $e^{-\kappa z}$, remaining localized within a distance of nanometers to micrometers from the interface. Z is the perpendicular axis and k represents the wavenumber related to the characteristic wavelength. In biological systems such boundaries are ubiquitous: they occur at membrane surfaces, between cytoplasmic and extracellular compartments, at myelin sheaths, and wherever dielectric discontinuities separate regions of differing ionic composition or polarizability. Evanescent modes arise from total internal reflection, waveguide-like boundaries, plasmonic or polaritonic coupling, dielectric discontinuities, and the molecular and ionic oscillations intrinsic to biological tissue.[41,42]

Because evanescent fields include longitudinal electric-field components which are absent in free-space transverse radiation, they enable energy exchange, information transfer, and phase-coherent coupling between different classes of electromagnetic excitations.[43] This is precisely the mechanism required to describe interface communication in biological systems, where near-field and far-field modes coexist and interact continuously.[44,45]

**5.3: Empirical Signatures of the Framework**

The MBW framework is not solely a mathematical taxonomy of biological scales: it makes physical claims about the nature of inter-scale coupling that are, in principle, distinguishable from those of standard volume-conduction and amplitude-based dynamical-systems models. Three predictions follow directly from the formal objects established in this paper and can guide experimental design to be proposed later. Each prediction identifies both what would constitute positive evidence for the framework and what result would require its revision. Some of the possible applications could provide great assistance to develop more efficient techniques for the study of neural behavior and to institute more sophisticated methods to rectify behavioral problems

1. **Activity-dependent evanescent decay constant at membrane interfaces.** If near-field evanescent coupling mediates signal transfer between the cellular sector \and the molecular sector, the electric-field amplitude measured perpendicular to a neuronal membrane should decay exponentially with distance z. This exponential form is itself predicted by standard near-field electrodynamics at any dielectric interface and is not unique to the MBW framework. The MBW framework predicts that $\kappa$ should shift systematically and reproducibly with neural activity state. Specifically, $\kappa$ (the material-dependent electrodynamic coupling coefficient) during an action potential should differ from $\kappa$ at rest by an amount predictable from the known change in membrane conductance ($\Delta G \sim 10–100\ nS/\mu m^2$), providing a parameter-constrained, quantitative prediction rather than a qualitative structural one. A second discriminating feature is spatial specificity: MBW predicts that the evanescent decay profile should be concentrated at membrane boundaries and not reproduced at intra-cellular locations of equivalent geometric permittivity contrast. Near-field scanning optical microscopy (NSOM) and plasmonic probe techniques,[46] capable of resolving sub-micron field structure at biological interfaces, provide the appropriate experimental platform. A measured activity-dependent shift in κ, correlated with action potential timing and consistent with the known ΔG of depolarization, would constitute evidence for physiologically coupled evanescent inter-sector modes rather than static near-field structure. A κ that is invariant across neural activity states, or that does not differ between membrane and intra-cellular measurement sites, would indicate that the near-field coupling mechanism is not modulated by the inter-sector dynamics the framework proposes.

2. **Longitudinal field component along the axonal axis.** The most physically distinctive claim of the MBW framework is that electromagnetic waves propagating through biological media acquire a longitudinal electric-field component absent in free-space transverse radiation. This is directly testable. In free space, a transverse electromagnetic wave carries no electric field component parallel to the direction of propagation. In a medium where longitudinal modes develop as the MBW framework predicts for neural tissue, a measurable E-field component parallel to the axonal propagation axis should be detectable during an action potential. Using a polarization-resolved near-field probe oriented parallel to the axon and positioned within the near-field decay length (∼1–10 μm from the membrane surface), any $E_{\parallel}$ signal above the noise floor constitutes evidence for the longitudinal mode. This experiment has not been performed in neural tissue and would constitute a decisive test of one of the framework's most novel physical predictions.[47]

3. **Nearest-neighbor temporal cascade across sectors.** The principle of biophysical scale locality states that each sector interacts directly only with its nearest neighbors in the hierarchy, with all long-range cross-scale effects emerging as higher-order indirect couplings through intervening sectors. This encodes a specific temporal prediction: a perturbation introduced at one sector level should produce detectable changes in adjacent sectors before it produces changes in non-adjacent sectors, with inter-sector propagation delays that scale with the number of intervening levels. Concretely, a targeted pharmacological disruption of microtubule vibrational modes is achievable with colchicine or nocodazole at doses that selectively destabilize microtubule dynamics[48] — should produce detectable changes in membrane potential fluctuations before producing changes in population-level EEG oscillatory power. Simultaneous patch-clamp recording (cellular resolution)[49] and high-density EEG (tissue-organ resolution)[50] during graded microtubule disruption would allow the inter-sector propagation delay to be measured directly.[51] If EEG changes are coincident with the molecular perturbation rather than delayed relative to cellular changes, the locality principle is violated and a non-tridiagonal coupling architecture must be considered.

### 6.0 Conclusion

This paper has established a conceptual wave-mechanics framework for multiscale biophysical communication in the brain. Beginning at the quantum scale, where molecular wavefunctions and quantized vibrational modes govern electrochemical behavior and extending through cellular, tissue, and organismal levels, we have described how electromagnetic signals are generated, modulated, and propagated as both transverse and longitudinal carrier waves through the neural network. Maxwell's equations provide the foundational formalism; QED provides a framework to describe the coupling of electromagnetic signals at the molecular and sub-atomic scale. Three empirical signatures are derived directly from the formal objects of this paper. Whereas the evanescent decay profile at membrane interfaces, the longitudinal field component along the axonal axis, and the nearest-neighbor temporal cascade across the near-term experimental tests of the framework.

**Acknowledgements:** The authors would like to thank prof. Dr. Ghazala Rehman CPpsychol, FBPsS (she/her), consultant Clinical Psychologist, University of Surrey, UK for some helpful discussions.

## References

1. Su, J; Song, Y.; Zhu, Z.; Huang, X.; Fan, J.; Qiao, J., Mao, F., "Cell-cell communication: new insights and clinical implications", Signal Transdu ct. Target Ther 2024, 9(1), 196. https://doi.org/10.1038/s41392-024-01888-z
2. Nevoit, G.; Poderiene, K.; Potyazhenko, M.; Mintser, O.; Jarusevicius, G., Vainoras, A., "The concept of biophotonic signaling in the human body and brain: rationale, problems and directions", Front Syst Neurosci 2025, 19, 1597329. https://doi.org/10.3389/fnsys.2025.1597329
3. Salles, A.; Bohn, K. M., Moss, C. F., "Auditory communication processing in bats: What we know and where to go", Behav Neurosci 2019, 133(3), 305-319. https://doi.org/10.1037/bne0000308

4. Wiltschko, W., Wiltschko, R., "Magnetic orientation and magnetoreception in birds and other animals", Journal of Comparative Physiology A 2005, 191(8), 675-693. https://doi.org/10.1007/s00359-005-0627-7
5. Schaworonkow, N., Nikulin, V. V., "Spatial neuronal synchronization and the waveform of oscillations: Implications for EEG and MEG", PLoS Comput Biol 2019, 15(5), e1007055. https://doi.org/10.1371/journal.pcbi.1007055
6. Nevoit, G.; Poderiene, K.; Potyazhenko, M.; Mintser, O.; Jarusevicius, G., Vainoras, A. The concept of biophotonic signaling in the human body and brain. Front. Syst. Neurosci. 2025, 19, 1597329. https://doi.org/10.3389/fnsys.2025.1597329
7. Pospíšil, P., Prasad, A., "Biofield in neural communication: A review and conceptual framework", Progress in Biophysics and Molecular Biology 2026, 200, 31-40. https://doi.org/10.1016/j.pbiomolbio.2026.02.005
8. Alexandrova, A. N. Quantum Mechanical Insights into Biological Processes at the Electronic Level. In Computational Modeling of Biological Systems; Dokholyan, N., Ed.; Springer: Boston, MA, 2012. https://doi.org/10.1007/978-1-4614-2146-7_6
9. Theise, N. D., Tuszynski, J. A., "Non-linearity, complexity, and quantization concepts in biology", Front Hum Neurosci 2025, 19, 1695510. https://doi.org/10.3389/fnhum.2025.1695510
10. Gassab, L.; Adams, B.; Hossen, Y. H. G.; Pusuluk, O.; Murugan, N. J.; Kominis, I. K.; Müstecaplıoğlu, Ö. E.; Petruccione, F., Craddock, T. J. A., "Quantum in Biology, Quantum for Biology, and Biology for Quantum: Mapping the Evidence and the Road Ahead", arXiv:2605.00205 [quant-ph] 2026. https://arxiv.org/abs/2605.00205
11. Pezzulo, G., Levin, M., "Top-down models in biology: explanation and control of complex living systems above the molecular level", Journal of The Royal Society Interface 2016, 13(124), 20160555. https://doi.org/10.1098/rsif.2016.0555
12. McCaig, C. D.; Rajnicek, A. M.; Song, B., Zhao, M., “Controlling cell behavior electrically: current views and future potential”, Physiol Rev 2005, 85(3), 943-978. http://doi.org/10.1152/physrev.00020.2004
13. Landau, L. D.; Lifshitz, E. M.; Pitaevskii, L. P. Electrodynamics of Continuous Media, 2nd ed.; Course of Theoretical Physics, Vol. 8; Pergamon Press: Oxford, 1984; ISBN 978-0-08-030275-1. https://www.sciencedirect.com/book/9780080302751/electrodynamics-of-continuous-media
14. Masood, S., Saleem, I., "Propagation of electromagnetic waves in extremely dense media", International Journal of Modern Physics A 2017, 32(15), 1750081. https://doi.org/10.1142/S0217751X17500816
15. Masood, S. (2025). Conceptual Approach to Quantum Electrodynamics and Applications, Institute of Physics Publishing. https://doi.org/10.1088/978-0-7503-6054-8
16. Masood, S. S. (1991). Photon mass in the classical limit of finite-temperature and-density QED. Physical Review D, 44(12), 3943. https://doi.org/10.1103/PhysRevD.44.3943

17. Liu, L.; Huang, B.; Lu, Y.; Zhao, Y.; Tang, X., Shi, Y., "Interactions between electromagnetic radiation and biological systems", iScience 2024, 27(3), 109201. https://doi.org/10.1016/j.isci.2024.109201
18. Cifra, M.; Fields, J. Z., Farhadi, A., "Electromagnetic cellular interactions", Progress in Biophysics and Molecular Biology 2011, 105(3), 223-246. https://doi.org/10.1016/j.pbiomolbio.2010.07.003
19. Buzsáki, G.; Anastassiou, C. A., Koch, C., "The origin of extracellular fields and currents — EEG, ECoG, LFP and spikes", Nature Reviews Neuroscience 2012, 13(6), 407-420. https://doi.org/10.1038/nrn3241
20. Fries, P., "Rhythms for Cognition: Communication through Coherence", Neuron 2015, 88(1), 220-235. https://doi.org/10.1016/j.neuron.2015.09.034
21. Schomer, D. L.; Lopes da Silva, F. H., Eds. Niedermeyer's Electroencephalography, 7th ed.; Oxford University Press: New York, 2018. http://doi.org/10.1093/med/9780190228484.001.0001
22. Wutz, A.; Melcher, D., Samaha, J., "Frequency modulation of neural oscillations according to visual task demands", Proceedings of the National Academy of Sciences 2018, 115(6), 1346-1351. https://doi.org/10.1073/pnas.1713318115
23. Bargmann, C. I., Marder, E., "From the connectome to brain function", Nat Methods 2013, 10(6), 483-490. http://doi.org/10.1038/nmeth.2451
24. Ladenbauer, J.; Augustin, M., Obermayer, K., "How adaptation currents change threshold, gain, and variability of neuronal spiking", Journal of Neurophysiology 2014, 111(5), 939-953. https://journals.physiology.org/doi/epdf/10.1152/jn.00586.2013
25. Hameroff, S.; Bandyopadhyay, A., Lauretta, D. S., "Microtubules are 'Fractal Time Crystals': Implications for Life and Consciousness", Journal of Consciousness Studies 2026, 33(1-2), 211-247. https://doi.org/10.53765/20512201.33.1.211
26. Mendoza-Halliday, D.; Major, A. J.; Lee, N.; Lichtenfeld, M. J.; Carlson, B.; Mitchell, B.; Meng, P. D.; Xiong, Y.; Westerberg, J. A.; Jia, X.; Johnston, K. D.; Selvanayagam, J.; Everling, S.; Maier, A.; Desimone, R.; Miller, E. K., Bastos, A. M., "A ubiquitous spectrolaminar motif of local field potential power across the primate cortex", Nature Neuroscience 2024, 27(3), 547-560. https://doi.org/10.1038/s41593-023-01554-7
27. Nunez, P. L., & Srinivasan, R. (2006). Electric Fields of the Brain: The Neurophysics of EEG. Oxford University Press. https://doi.org/10.1093/acprof:oso/9780195050387.001.0001
28. Hameroff, S., "'Orch OR' is the most complete, and most easily falsifiable theory of consciousness", Cognitive Neuroscience 2020, 12, 1-3. http://doi.org/10.1080/17588928.2020.1839037
29. Mohsin, M.; Cantiello, H. F.; Cantero, M. d. R.; Marucho, M. "Electrical Oscillations in Microtubules." Scientific Reports 2025, 15, 41106. https://doi.org/10.1038/s41598-025-24920-w

30. Pokorný, J.; Pokorný, J., Vrba, J., "Generation of Electromagnetic Field by Microtubules", Int J Mol Sci 2021, 22(15), 8215. http://doi.org/10.3390/ijms22158215
31. Hameroff, S., Penrose, R., "Consciousness in the universe: A review of the 'Orch OR' theory", Physics of Life Reviews 2014, 11(1), 39-78. https://doi.org/10.1016/j.plrev.2013.08.002
32. Keppler, J., "Macroscopic quantum effects in the brain: new insights into the fundamental principle underlying conscious processes", Frontiers in Human Neuroscience 2025, 19, 1676585. http://doi.org/10.3389/fnhum.2025.1676585
33. Masood, S. S., "QED plasma in the early universe", Arabian Journal of Mathematics 2019, 8(3), 183-192. https://doi.org/10.1007/s40065-018-0232-6

34. Angelsky, O. V.; Zenkova, C. Y.; Hanson, S. G., Zheng, J., "Extraordinary Manifestation of Evanescent Wave in Biomedical Application", Frontiers in Physics 2020, Volume 8 - 2020. https://www.frontiersin.org/journals/physics/articles/10.3389/fphy.2020.00159
35. Jackson, Classical Electrodynamics, 3rd ed. 1998, see Sec. 7.6: "For total internal reflection, the transmitted wave has an imaginary normal component of the wave vector, and the fields are no longer purely transverse." https://www.wiley.com/en-us/shop/general-physics/classical-electrodynamics-3rd-edition-p-9780471309321
36. Voytek, B.; Kramer, M. A.; Case, J.; Lepage, K. Q.; Tempesta, Z. R.; Knight, R. T., Gazzaley, A., "Age-Related Changes in 1/f Neural Electrophysiological Noise", The Journal of Neuroscience 2015, 35(38), 13257-13265. http://doi.org/10.1523/jneurosci.2332-14.2015
37. Tufail, Y.; Matyushov, A.; Baldwin, N.; Tauchmann, M. L.; Georges, J.; Yoshihiro, A.; Helms Tillery, S. I.; Tyler, W. J. "Transcranial Pulsed Ultrasound Stimulates Intact Brain Circuits." Neuron 2010, 66(5), 681–694. https://doi.org/10.1016/j.neuron.2010.05.008
38. Dell'Italia, J.; Sanguinetti, J. L.; Monti, M. M.; Bystritsky, A., Reggente, N., "Current State of Potential Mechanisms Supporting Low Intensity Focused Ultrasound for Neuromodulation", Frontiers in Human Neuroscience 2022, 16:872639 https://doi.org/10.3389/fnhum.2022.872639
39. Pokorný, J.; Vedruccio, C.; Cifra, M., Kučera, O., "Cancer physics: diagnostics based on damped cellular elastoelectrical vibrations in microtubules", Eur Biophys J 2011, 40(6), 747-759. http://doi.org/10.1007/s00249-011-0688-1
40. Bertolotti, M.; Sibilia, C.; Guzman, A. Evanescent Waves. In Evanescent Waves in Optics. Springer Series in Optical Sciences, Vol. 206; Springer: Cham, 2017. https://doi.org/10.1007/978-3-319-61261-4_2
41. Bliokh, K. Y.; Bekshaev, A. Y., Nori, F., "Extraordinary momentum and spin in evanescent waves", Nature Communications 2014, 5(1), 3300. https://www.nature.com/articles/ncomms4300.pdf
42. Zhang, Z., Ahn, P., Dong, B., Balogun, O., & Sun, C., "Quantitative Imaging of Rapidly Decaying Evanescent Fields Using Plasmonic Near Field Scanning Optical Microscopy."

Scientific Reports 2013, 3, 2803. https://pmc.ncbi.nlm.nih.gov/articles/PMC3786296/pdf/srep02803.pdf

43. Butt, M. A., "Dielectric Waveguide-Based Sensors with Enhanced Evanescent Field: Unveiling the Dynamic Interaction with the Ambient Medium for Biosensing and Gas-Sensing Applications—A Review", Photonics 2024, 11(3), 198. https://www.mdpi.com/2304-6732/11/3/198
44. Angelsky, O. V.; Zenkova, C. Y.; Hanson, S. G., Zheng, J., "Extraordinary Manifestation of Evanescent Wave in Biomedical Application", Frontiers in Physics 2020, Volume 8 - 2020. https://www.frontiersin.org/journals/physics/articles/10.3389/fphy.2020.00159
45. Jackson, J. D. Classical Electrodynamics, 3rd ed.; Wiley: New York, 1998; Sections 7.7 (Fields in Total Internal Reflection; Evanescent Waves), pp. 307–314. https://www.wiley.com/en-us/shop/general-physics/classical-electrodynamics-3rd-edition-p-9780471309321
46. Betzig, E.; Trautman, J. K., "Near Field Optics: Microscopy, Spectroscopy, and Surface Plasmonics", Science 1992, 257, 189–195. https://doi.org/10.1126/science.257.5067.189
47. Electromagnetic waves in cylindrical geometries generically develop longitudinal electric-field components, even in classical Maxwell theory (Niziev, V. G., Nesterov, A. V., "Longitudinal fields in cylindrical and spherical modes", Journal of Optics A: Pure and Applied Optics 2008, 10(8), 085005. https://doi.org/10.1088/1464-4258/10/8/085005). Therefore, the presence of a nonzero $E_\parallel$ near an axon is not itself diagnostic. The decisive test for the MBW framework is whether the longitudinal component observed during an action potential matches the magnitude, spatial structure, and dispersion predicted by the MBW decomposition (eq. 6), rather than the longitudinal fields expected from standard cylindrical-mode electrodynamics.
48. Jordan, M. A.; Wilson, L., "Microtubules as a Target for Anticancer Drugs", Nat. Rev. Cancer 2004, 4, 253–265. https://doi.org/10.1038/nrc1317
49. Hamill, O. P.; Marty, A.; Neher, E.; Sakmann, B.; Sigworth, F. J., "Improved Patch Clamp Techniques for High Resolution Current Recording from Cells and Cell Free Membrane Patches", Pflügers Arch. 1981, 391, 85–100. https://doi.org/10.1007/BF00656997
50. Michel, C. M.; Brunet, D., "High Resolution EEG: Electrical Neuroimaging", Clin. Neurophysiol. 2019, 130, 1680–1697. https://doi.org/10.1016/j.clinph.2019.01.001
51. Breakspear, M., "Dynamic Models of Large Scale Brain Activity", Nat. Neurosci. 2017, 20, 340–352. https://doi.org/10.1038/nn.4497